\documentclass[twocolumn]{article}

\usepackage{graphicx}
\usepackage{amsmath}
\usepackage{mathrsfs}
\usepackage{amssymb}
\usepackage{cite}
\usepackage[labelfont={bf,sf},footnotesize,indention=0cm]{caption}

\begin{document}

\title{XUV digital in-line holography using high-order harmonics}

\author{G. Genoud, O. Guilbaud, E. Mengotti, S.-G. Pettersson, 
E. Georgiadou, \\ E. Pourtal, C.-G. Wahlstr\"{o}m and A. L'Huillier}

\date{}

\maketitle

\begin{abstract}
A step towards a successful implementation of time-resolved digital in-line holography with extreme ultraviolet radiation is presented. Ultrashort  XUV pulses are produced as high-order harmonics of a femtosecond laser and a Schwarzschild objective is used to focus harmonic radiation at 38 nm and to produce a strongly divergent reference beam for holographic recording. Experimental holograms of thin wires are recorded and the objects reconstructed. Descriptions of the simulation and reconstruction theory and algorithms are also given. Spatial resolution of few hundreds of nm is potentially achievable, and micrometer resolution range is demonstrated. 
\end{abstract}

\section{Introduction}

Advanced coherent and short-pulse photon sources in the XUV and X-ray range are being extensively developed with the goal to provide both temporal and spatial resolution, thus allowing scientists to explore at the same time structure and dynamics of matter. Examples of this development are the effort to build X-ray free electron lasers at various places in the world \cite{FEL} and to develop seeded soft-x-ray lasers \cite{ZeitounNature2004}. One of the first experiments performed at FLASH, XUV-free electron laser facility in Hamburg is time-resolved X-ray holography using a clever single-shot design \cite{Chapman}. Much smaller sources are those provided by high-order harmonic generation in gases \cite{Handbook}. These sources have the advantage of being table-top and affordable in university laboratories. The temporal resolution reaches the attosecond time scale, while the wavelength range varies from the UV to the soft X-ray spectral regions. The highest conversion efficiencies, of the order of a few times $10^{-5}$, and energy per pulse (fractions of $\mu$J) are obtained when Xe or Ar are used for generating the harmonics, with wavelengths typically between 30 and 50 nm. The spatial coherence has been measured to be excellent, comparable to that of the generating laser beam \cite{BartelsScience2002,LederoffPRA2000}.
 
These properties make this radiation an interesting candidate for time-resolved microscopy \cite{Rocca}, in particular using holographic approaches \cite{Gabor}, at least in the long-wavelength part of the XUV range (around 30 nm) where the number of photons per pulse and per harmonic can be as high as $10^9-10^{10}$. A few promising experiments have been performed, using, in particular, digital holographic techniques in the XUV \cite{BartelsScience2002,Morlens2006} and X-ray range \cite{Jacobsen1990,McNulty1994,Eisebitt2004,Tobey}. Bartels \textit{et al.} \cite{BartelsScience2002} used a very simple setup, where an object was placed directly in a harmonic beam, containing harmonics of orders 17 through 23.
The resulting interference patterns, produced by the diverging harmonic beam, acting as a reference wave, and the waves diffracted from the object, were recorded by an X-ray CCD camera. The resolution
achieved was not very high due to a low numerical aperture, and the use of several harmonics blurred the image. However, images of the tip of a scanning microscope could be reconstructed. Morlens \textit{et al.} \cite{Morlens2006} used two multilayer spherical mirrors to focus the beam on the sample in order to achieve a larger numerical aperture and to select only one harmonic, the $25^{\textrm{th}}$. Holograms of spider threads were recorded and reconstructions with resolution below one micrometer could be successfully performed. In both cases, however, acquisitions of over 1000 shots were needed to achieve high signal/noise ratio. Recently Tobey \textit{et al.} \cite{Tobey} performed time-resolved holography using high-order harmonics. The experiment was done with a simple single-reflection geometry and without any focusing optics. In the present work, we use high-order harmonics produced by a powerful 10 Hz, 35 fs, titanium-sapphire laser system to perform digital in-line holography of micrometer wires at 38 nm wavelength, using focusing optics with higher spatial resolution than in the previous experiments. High quality holograms are obtained by adding about 100 images. Single shot images are also obtained, though with low signal to noise ratio. Images are successfully reconstructed by using a reconstruction algorithm. In Section 2 we describe the experimental method and results, while in Section 3, we give details of our reconstruction algorithms and apply them to the analysis of the experimental holograms. In Section 4, we discuss how to improve the experiments to realise high-quality single shot digital in-line holography, leading the way to time-resolved holographic microscopy in the femtosecond regime.

\section{Experiment}

\begin{figure}
\begin{center}
\includegraphics[width=8cm]{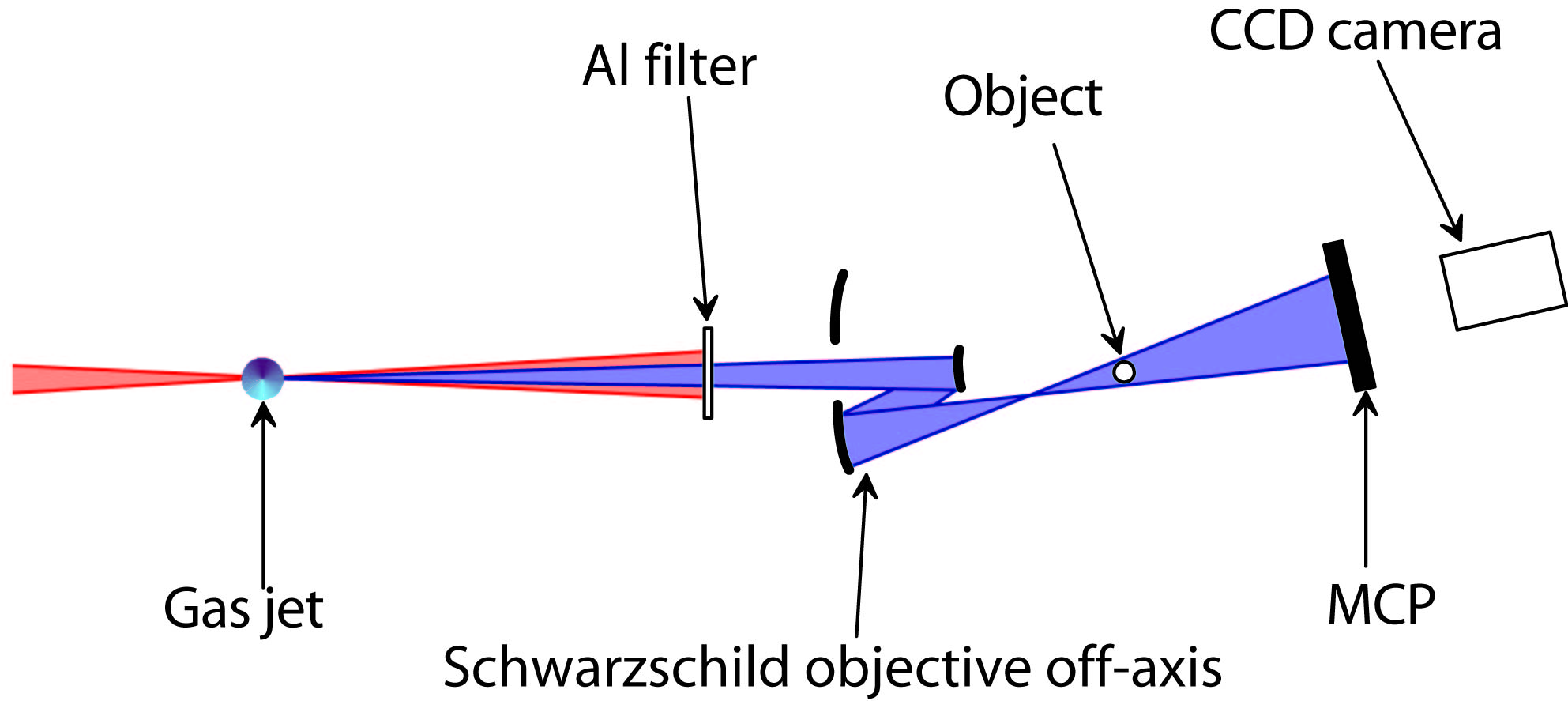}
\caption{Experimental setup. The laser beam is focused into a gas jet producing harmonics, which then pass trough an aluminium filter. A multi-layer coated Schwarzschild objective focus the harmonic beam on an object and the resulting interference pattern is recorded by a multi-channel plate and a CCD camera.} \label{setup}
\end{center}
\end{figure}

\begin{figure}
\begin{center}
\includegraphics[width=8cm]{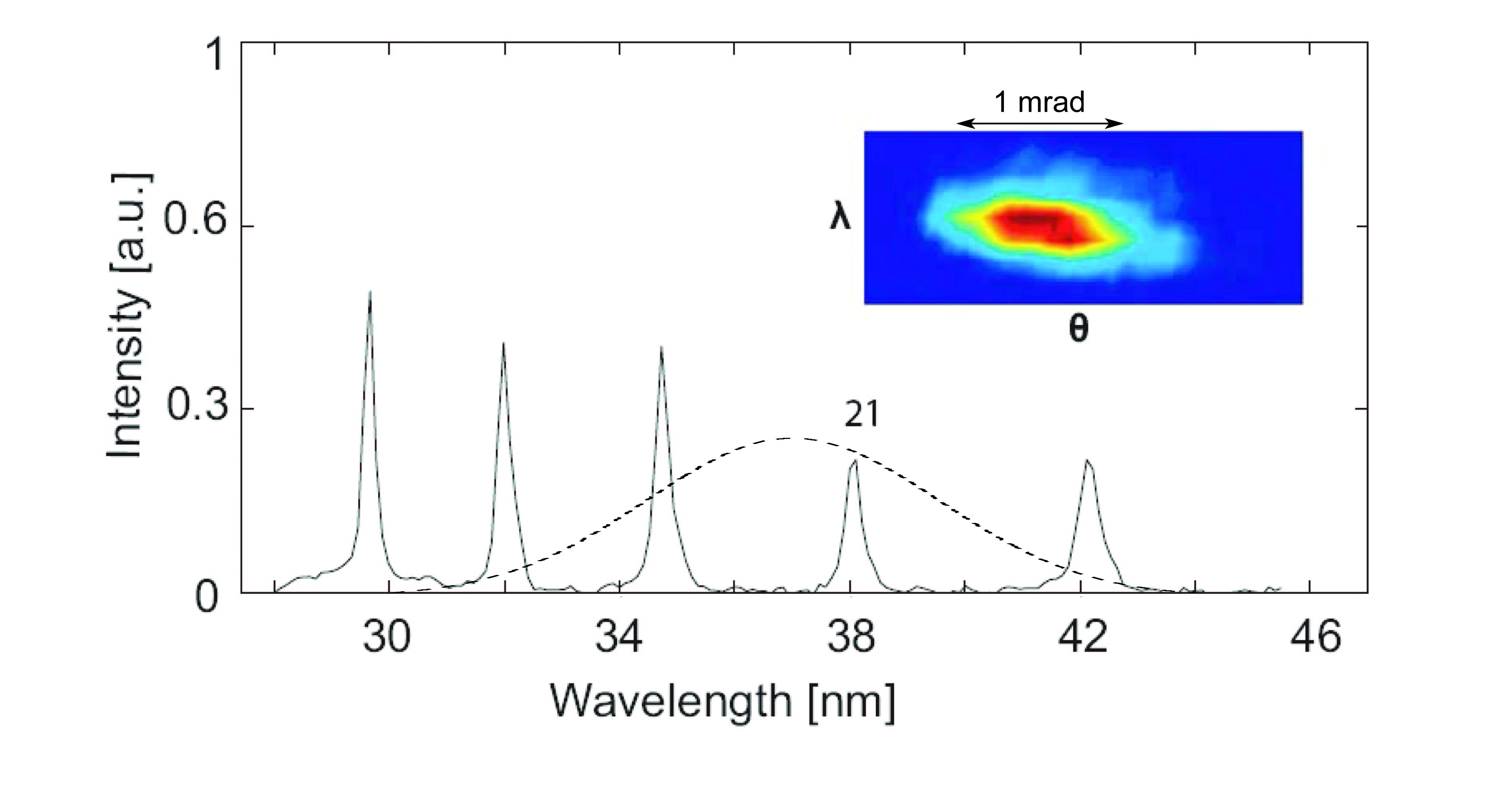}
\end{center}
\caption {High harmonic spectrum and profile of the 21$^{\textrm{st}}$ harmonic. The dashed line represents the transmission of the Schwarzschild objective.} \label{spectrum}
\end{figure}

Fig. \ref{setup} describes our experimental setup. The laser beam
is a low-energy arm of the 40 TW laser system at the Lund Laser Center, a titanium-sapphire laser based on the chirped-pulse
amplification technique. Pulses are delivered at a 10 Hz repetition
rate, 35 fs duration, 800 nm central wavelength and up to 100 mJ energy. 10 mJ are actually used to generate harmonics by focusing loosely (1 m focal length) the laser beam in an argon gas jet. After that the fundamental beam is blocked by a 200 nm-thick aluminium filter. Fig. \ref{spectrum} shows a typical harmonic spectrum recorded with a flat-field XUV spectrometer. A multi-layer coated Schwarzschild objective is used to focus tightly the harmonic beam and at the same time to select a narrow spectral region. An object (wire) is located a few mm after the image point of the harmonic source and  holograms are recorded by using imaging multichannel plates (MCP) coupled to a phosphor screen. A CCD camera is used to record the image on the phosphor screen. 

A Schwarzschild objective is an optical system consisting of two
spherical mirrors, one convex and one concave of which the centers of
curvature coincide. A tight focusing can be obtained. In our design the two radii of curvature are equal to 
3.39 and 9.17 cm respectively, with a resulting focal length of 2.69 cm. The Schwarzschild objective is placed 200 cm from the gas jet producing the harmonics and the detector 30 cm further away. The object is located about 3 mm after the focus of the Schwarzschild objective. To avoid loosing significant XUV energy, we use it in a slightly off-axis 
geometry as shown in Fig. \ref{setup}, without introducing significant aberration. Due to this off-axis configuration, the beam going out of the objective is forming an angle of 10$^{\circ}$ with the ingoing beam. It is coated with a multilayer structure giving a transmission window centered at 
$\lambda=37 \pm 3$ nm. The maximum transmission is approximately 40 \%.

The wavelength transmission of the objective allows us to select the 21$^\textrm{{st}}$ harmonic. The inset at the top of  Fig. \ref{spectrum} represents the far-field profile of that harmonic, with a 1 mrad divergence. The energy available in the 21$^\textrm{{st}}$ harmonic is measured to be 5 nJ which corresponds to 10$^9$ photons per pulse, reduced by a factor of 5 after transmission through the Al filter.

\begin{figure}
\begin{center}
\includegraphics[width=8cm]{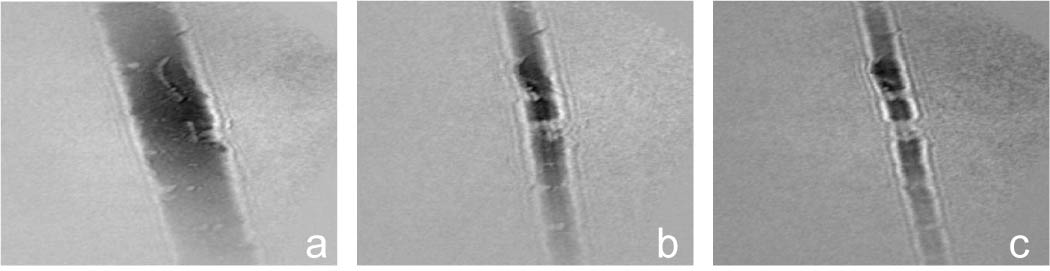}
\end{center}
\caption {Holograms recorded after 200 shots with different objects: a hair of 65
\textrm{$\mu$}m diameter (a), a wire  of 18
\textrm{$\mu$}m diameter (b) and a spider thread of 5
\textrm{$\mu$}m diameter (c).} \label{holograms}
\end{figure}

The theoretical resolution of the experiment is given by:
\begin{equation}
R=\frac{\lambda}{2 tan(\theta)}=\frac{\lambda L}{d},
\label{equ_aperture}
\end{equation}
where $\theta$ is the half angle of 
aperture, $\lambda$ the wavelength
of the radiation used, $L$ the distance between the object and the detector and $d$ the diameter of the detector area illuminated by the beam. In our setup, the half angle of 
aperture after the Schwarzschild objective is 4.3$^{\circ}$, leading to a resolution of 250 nm. Recorded holograms of wires of diameters varying from a few tens to a few micrometers are shown in Fig. \ref{holograms}. These results are obtained after integration of 200 images.

\section{Reconstruction}

In this Section the algorithm used for simulation of holograms and reconstruction of experimental images is presented. Many innovative techniques have been put together to allow reconstruction of experimental datas. Due to the tight focusing of the beam on the object, the beam has gaussian properties, that have been taken into account. Aliasing is also avoided in this process. Furthermore the problem of the twin images is solved by introducing an iteration process.

\begin{figure}
\begin{center}
\includegraphics[width=8.4cm]{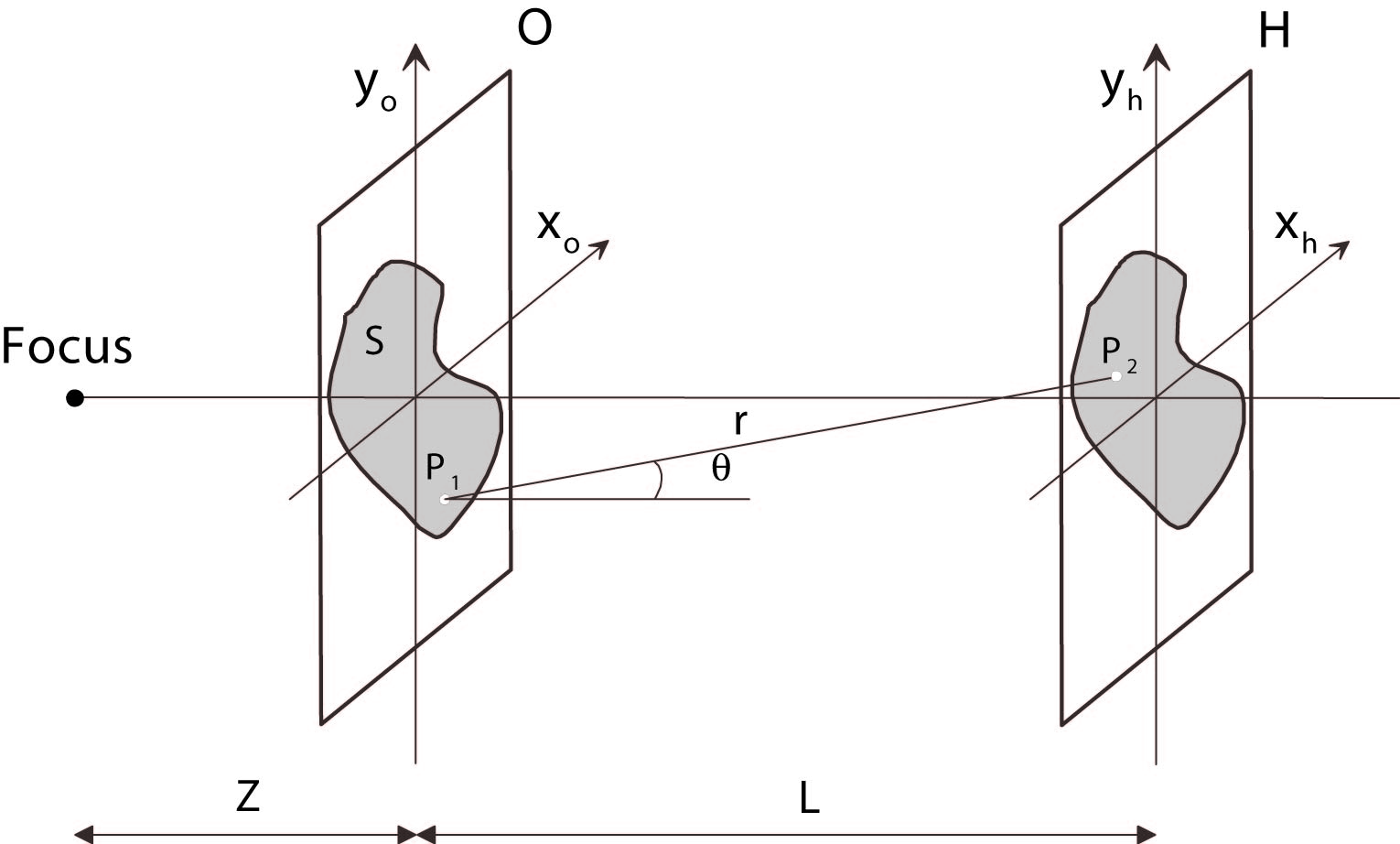}
\caption{Illustration of the theoretical formalism and notations. Two Fresnel planes, the object plane ($O$) and the hologram plane ($H$), are indicated.}
\label{fresnel_planes}
\end{center}
\end{figure}

Our reconstruction algorithm is based on the Fresnel-Kirchoff integral, illustrated in Fig. \ref{fresnel_planes} \cite{Schnars2004}. Every point in the object plane $O$ ($x_o$, $y_o$) is the source of a spherical wavelet. Their addition in the hologram plane $H$ ($x_h$, $y_h$) is the electromagnetic field :
\begin{eqnarray}
E_H(x_h,y_h,Z+L)&=&\frac{i}{\lambda}\int_S E_O(x_o,y_o,Z)
\frac{e^{-ikr}}{r} \nonumber \\
&  & \times \cos\theta dx_ody_o, \label{fresnel}
\end{eqnarray}
where $r$ is the distance from the point $P_1$ to the point $P_2$, $\theta$ is the angle between $P_1P_2$ and the normal to the planes, $S$ is the area over which the integral is carried out. $Z$ and $L$ define the position of the object and hologram planes as illustrated in Fig. \ref{fresnel_planes}.

In the paraxial approximation, $|x_h-x_o|$, $|y_h-y_o| \ll L$,  $\cos\theta \simeq$1, $r$ can be expressed to first order as:
\begin{equation}
r \approx
L+\frac{(x_o^2+y_o^2)}{2L}+\frac{(x_h^2+y_h^2)}{2L}-\frac{(x_o
x_h+y_o y_h)}{L}.
\end{equation}
Using $F_O^L (x_o,y_o,Z)= E_O(x_o,y_o,Z) exp(-ik\frac{x_o^2+y_o^2}{2L})$ and its Fourier transform $\widetilde{F}_O^L$, the resulting field can be written as 
\begin{eqnarray}
\lefteqn{E_H(x_h,y_h,Z+L)=} \nonumber \\
& &  \frac{ie^{-ikL}}{\lambda L }  e^{-ik\frac{x_h^2+y_h^2}{2L}}  \widetilde{F}_O^L
(\frac{k x_h}{2\pi L}, \frac{k y_h}{2\pi L},Z). \label{equ_EH1}
\end{eqnarray}

\begin{figure}
\begin{center}
\includegraphics[width=8cm]{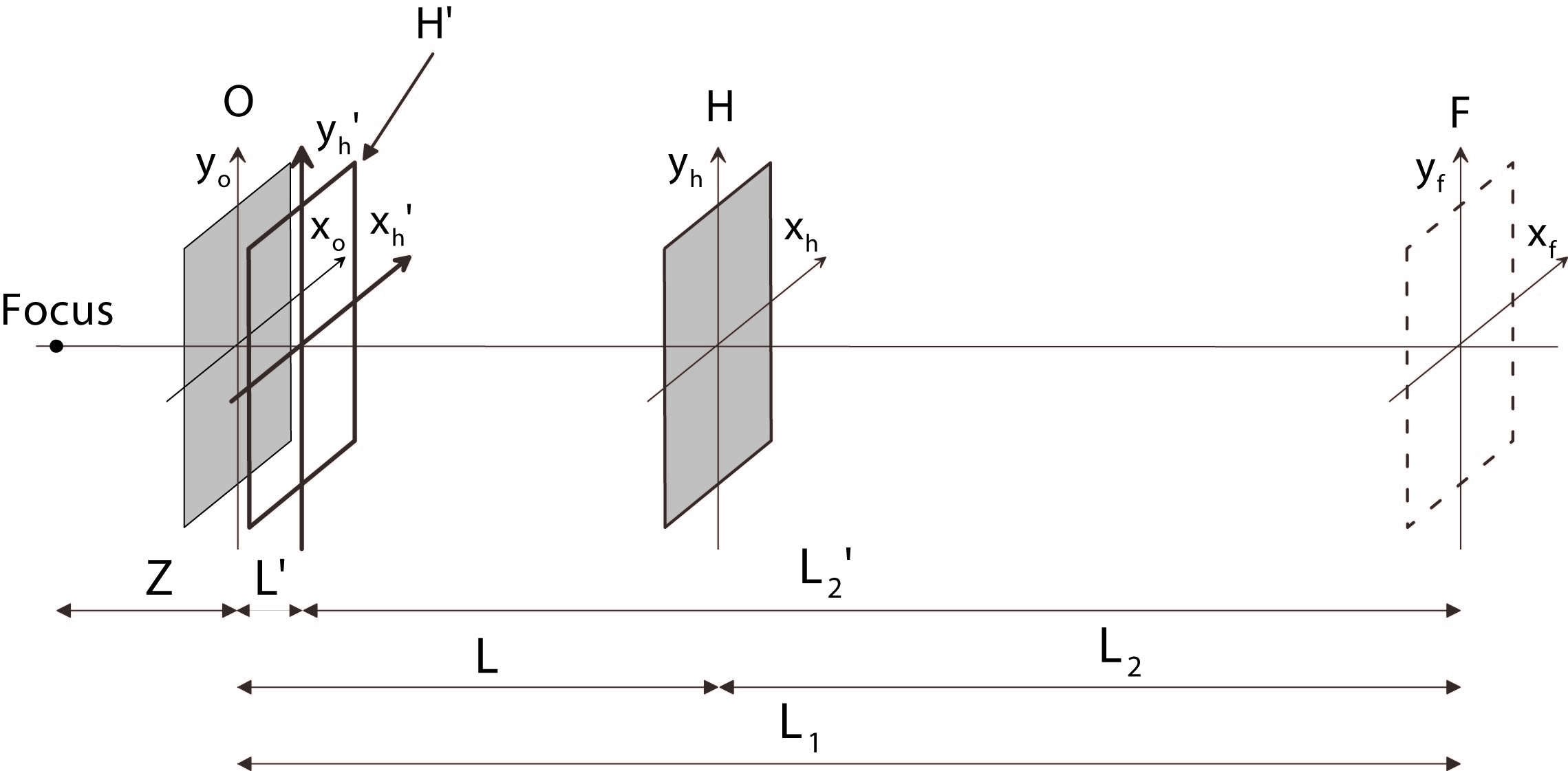}
\caption{Far-field, object and hologram planes, $F$, $O$ and $H$, respectively. A new hologram plane ($H'$) closer to the object plane is also introduced to solve the problem of the pixel size (see text).}
\label{reconstruction_planes}
\end{center}
\end{figure}

We now consider a third plane, ($F$), called far-field plane, far
from the object and hologram planes, as shown
in Fig. \ref{reconstruction_planes}. The electric field in the far-field plane can be determined from the electric field in any of the two other planes \cite{Deng2000}, as
\begin{eqnarray}
\lefteqn{E_F(x_f,y_f,Z+L_1)=} \nonumber \\
& & \frac{ie^{-ikL_1}}{\lambda L_1}  e^{-ik \frac{x_f^2+y_f^2}{2L_1}} \widetilde{F}_O^{L_1}(\frac{kx_f}{2\pi L_1},\frac{ky_f}{2\pi L_1},Z)= \nonumber \\
 & &  \mbox{}  \frac{ie^{-ikL_2}}{\lambda L_2} e^{-ik \frac{x_f^2+y_f^2}{2L_2}}  \widetilde{F}_H^{L_2}(\frac{kx_f}{2\pi L_2},\frac{ky_f}{2\pi L_2},Z+L).
\end{eqnarray}
$L_1$ and $L_2$ are the distances between $F$ and $O$, and between $F$ and $H$, respectively (see Fig. \ref{reconstruction_planes}). By dividing the two equations 
an expression for $\widetilde{F}_H^{L_2}$ as a function of $\widetilde{F}_O^{L_1}$ is obtained, and vice versa.  

To simulate a hologram we can then calculate
\begin{eqnarray}
\lefteqn{E_H(x_h,y_h,Z+L)=e^{ik \frac{x_h^2+y_h^2}{2L_2}} F_H^{L_2}(x_h,y_h,Z+L)} \nonumber \\
& = & e^{ik \frac {x_h^2 +y_h^2}{2L_2}} FT^{-1}[G\widetilde{F}_O^{L_1}(\frac{kx_f}{2\pi
L_1},\frac{ky_f}{2\pi L_1},Z)], \label{equ_eh}
\end{eqnarray}
where 
\begin{equation}
G = \frac{L_2}{L_1}e^{-ik (L_1-L_2)}e^{-ik
\frac{x_f^2+y_f^2}{2}(\frac{1}{L_1}-\frac{1}{L_2})}. \label{equ_F}
\end{equation}
Similarly, to reconstruct the object, we use
\begin{eqnarray}
\lefteqn{E_O(x_o,y_o,Z)=e^{ik\frac {x_o^2 +y_o^2}{2L_1}}} \nonumber \\
&  \times & FT^{-1}[G' \widetilde{F}_H^{L_2}(\frac{kx_f}{2\pi
L_2},\frac{ky_f}{2\pi L_2},Z+L)], \label{equ_eo}
\end{eqnarray}
where
\begin{equation}
G'= \frac{L_1}{L_2} e^{-ik(L_2-L_1)}  e^{-ik
\frac{x_f^2+y_f^2}{2}(\frac{1}{L_2}-\frac{1}{L_1})}.
\end{equation}

This reconstruction method works quite well for plane or collimated waves. 
However in our case, tight focusing optics is 
placed before the object, leading to a strong beam divergence. The
beam cannot be approximated as a plane wave anymore. Moreover, 
the typical size of the object is orders of
magnitude smaller than the size of the detector and the pixel sizes in the
hologram and far-field planes are large. The phase
factors of different terms of $E_H$ and $E_O$ therefore vary very rapidly, 
leading to aliasing and picture
replicas \cite{Allebach}. To overcome this problem, we introduce
a new hologram plane, hologram plane ($H'$) (see
Fig.~\ref {reconstruction_planes}). This plane is situated close to
the object plane and there is almost no magnification
between the two planes so that the algorithm described above can be
applied. A simple transformation is then made to obtain the
hologram in $H$.
\underline{}
More formally, the field in the object plane can be written as 
\begin{equation}
E_O(x_o,y_o,Z)= \overline{E}_O (x_o,y_o,Z) exp{(-i k \frac{x_o^2+y_o^2}{2Z})},
\end{equation}
where $\overline{E}_O$ is a slowly varying term (product of the object transmission field and 
the slowly varying part of the reference field, taken to be Gaussian) while the phase term is a rapidly varying spherical term. We apply the following coordinate reduction:
\begin{equation}
(x_h,y_h)\mapsto(x_{h'},y_{h'})=(\frac{L'}{L} x_h, \frac{L'}{L} y_h),
\label{equ_coord}
\end{equation}
with $L'=LZ/(L+Z)$. After some elementary mathematical manipulation, the field in the hologram plane can be expressed as:
\begin{eqnarray}
\lefteqn{E_H(x_h,y_h,Z+L) = \frac{Z-L'}{Z}e^{-ik(L-L')} } \nonumber \\
& \times & e^{-ik (x_{h'}^{2}+y_{h'}^2)\frac{L-L'}{2L'^2}}  \overline{E}_{H'}(x_{h'},y_{h'},Z+L').\label{eq_lp}
\end{eqnarray}
where $\overline{E}_{H'}(x_{h'},y_{h'},Z+L')$ is the field in $H'$ corresponding to the slowly-varying field  $\overline{E}_O$ in the object plane. The terms in front $\overline{E}_{H'}$ in Eq.~(\ref{eq_lp}) are, apart from a constant factor,
pure phase terms that vanish when calculating the hologram intensity.
This equation can be interpreted as follows: the diffraction
calculation with a very divergent reference wave is equivalent to
the diffraction calculation with a plane wave, on a plane
$H'$ separated from the object by $L'$, followed
by a magnification by $(L+Z)/Z$.
The simulation algorithms described before can now be applied on $\overline{E}_O$, using $H'$.
To get the real hologram in $H$ we just make a scale
change using Eq.~(\ref{equ_coord}). The reconstruction algorithm is based on the same idea.

In the reconstruction process, the phase information is lost because
only the intensity of the hologram is recorded. To retrieve this
information, an iteration process inspired from the Gerchberg-Saxton algorithm, similar to the one described in
\cite{Spence1999}, has been implemented. The aim of this process is
to eliminate the twin images following reconstruction in in-line
holography. Here are the different steps:
\begin{enumerate}
\item  Reconstruction of the object,
using the intensity information of the hologram (no phase
information).
\item Simulation of a new hologram using
the real part of the reconstructed object as object field.
\item Replacement of the intensity of this new hologram by the intensity
of the original hologram. The phase of the simulated hologram is
now used as a first guess of the phase of the field on the hologram
plane.
\item Reconstruction of the object using this last field guess.
\item Repeat steps 2 through 4 until convergence.
\end{enumerate}

\begin{figure}
\begin{center}
\includegraphics[width=8cm]{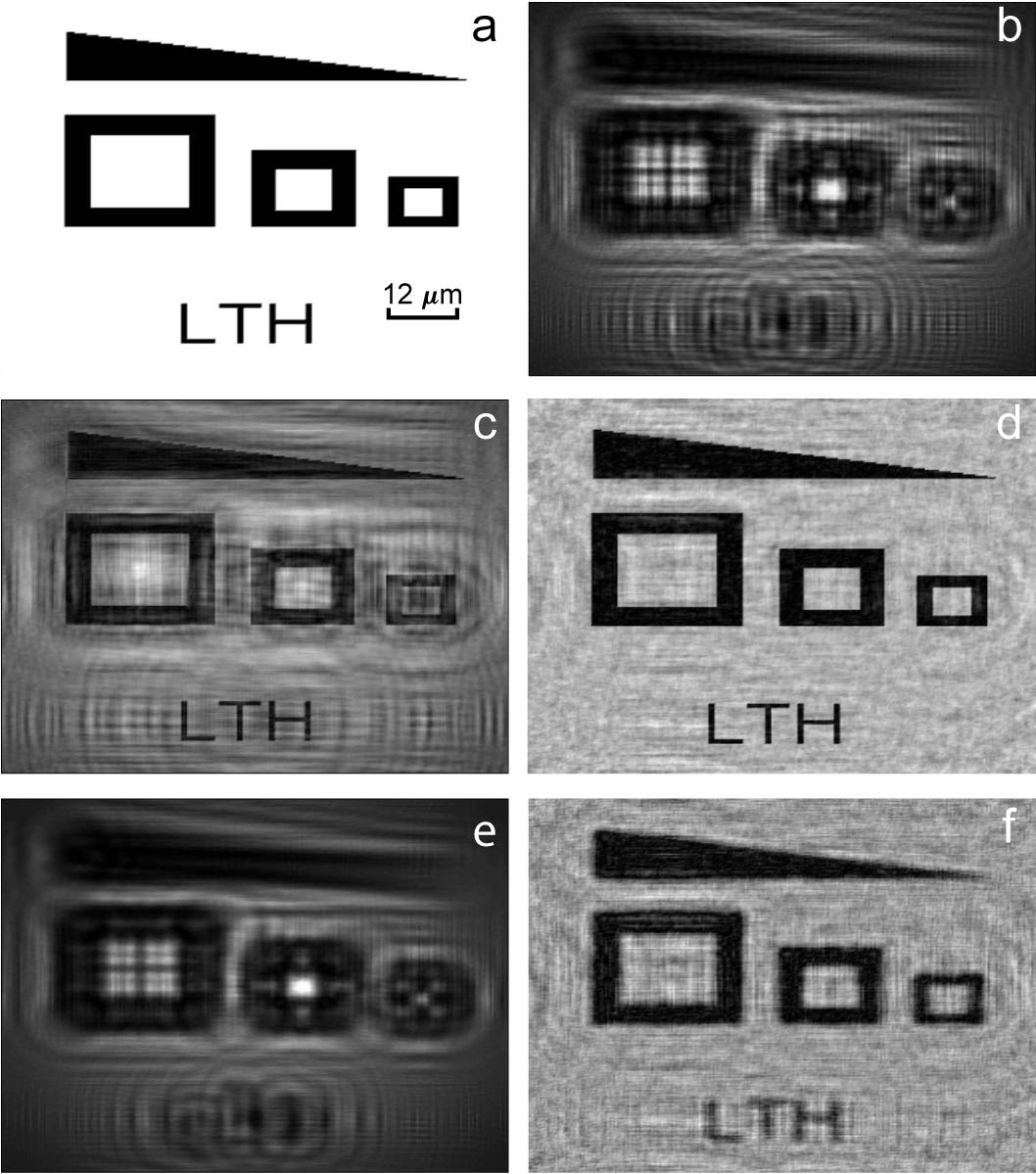}
\caption{A hologram is simulated in (b) from the object in (a) with one harmonic. The object is reconstructed in without (c) and with iterations (d). In (e) the hologram is simulated with three harmonics and reconstructed in (f). The holograms are simulated and reconstructed with the
experimental conditions: $L=0.27$ m; $Z=1$ mm; $pix_{hologram}=95$ $\mu$m; $L_1=10^6$ m.
The object is illuminated by a Gaussian beam with a half angle of aperture of 4.3 $^{\circ}$.}
\label{simulation}
\end{center}
\end{figure}

Fig. \ref{simulation} 
illustrates the performance of this algorithm on simulated data. In (a) the object 
is represented and in (b) the simulated hologram. The object is reconstructed 
in (c) without iteration and in (d) with 100 iterations. The iteration process clearly improves the quality of the reconstruction. We also studied the simulation and reconstruction of an hologram generated by many harmonics. In (e) the hologram is simulated with three harmonics (the 19$^{\textrm{th}}$, the 21$^{\textrm{st}}$ and the 23$^{\textrm{rd}}$). Despite the relative blurring of the fringes in the hologram, the object can still be reconstructed in (f) also with 100 iterations.

In the experiment, the images are recorded by a digital camera and processed by a computer program. 
Thus the influence of the hologram pixel size has to be taken
into account to estimate the experimental resolution, which will depend on
the pixel size of the reconstructed object given by the algorithm.
As seen in Eq. (\ref{equ_eh}) and (\ref{equ_eo}) we need to calculate
numerically several 2D Fourier transforms and their inverses to
simulate holograms or reconstruct object from holograms. These
calculations involve a discretization of $H$ and $O$ in arrays of pixels. Understanding the relations between
these pixel sizes is crucial. The size of the reconstructed pixel of
$O$ has indeed to be of the order of the theoretical
optical resolution in order to reach the full power of the method.  
Numerically, a reconstructed object
with pixel size smaller than the optical resolution will exhibit
diffraction blurring of the order of the optical resolution. For simplicity we 
consider the shape of the pixels of the object and
of the hologram as squares. 

In the case of Fresnel diffraction calculated with a 2D Fast Fourier
Transform, the pixel sizes in $F$ and in $O$ are related by the simple expression:

\begin{equation}
pix_{far field}=\frac{\lambda L_1}{N} \frac{1}{pix_{object}},
\label{pix_size1}
\end{equation}
where $L_1$ is the distance between the far field plane and the
object, $N$ is the number of pixels in one raw of the object plane,
$pix_{far field}$ is the length of the pixel side in $F$ and
$pix_{object}$ is the length of the pixel side in $O$. A similar expression 
relates the pixel size in $H$ and in $F$, so that 
$pix_{hologram}=pix_{object} L_2/L_1$. Taking
into account the reduction factor introduced before, we have 
\begin{eqnarray}
pix_{object}& =&\frac{L_1 Z}{L_1(L+Z)-LZ}pix_{hologram}  \nonumber \\
&\stackrel{L_1 \to \infty}{\longrightarrow}&\frac{Z}{L+Z} pix_{hologram}. \label{equ_resolution}
\end{eqnarray}
The pixel size of the reconstructed object depends on
$L_1$, $L$ and $Z$ and on the pixel size of the hologram. The
minimum value is reached when $L_1 \to \infty$.
To obtain an even smaller pixel size, we should thus reduce the pixel size
of the hologram by interpolating the experimental hologram or
recording with a device with smaller pixels. Note however that, the resolution of the experiment is not limited by the reconstruction algorithm but only by the optical resolution given by Eq. (\ref{equ_aperture}).

\begin{figure}
\begin{center}
\includegraphics[width=8cm]{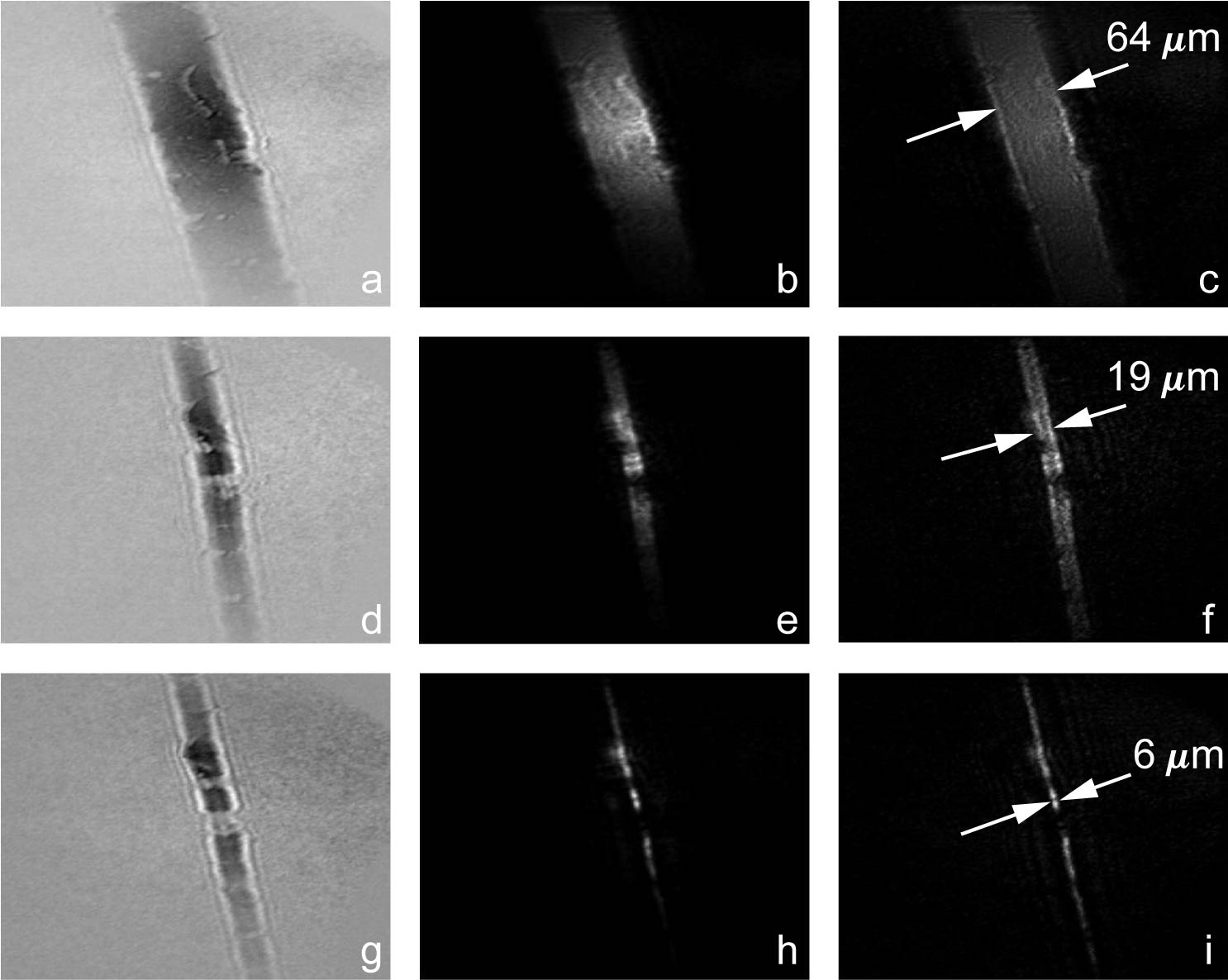}
\end{center}
\caption {Holograms recorded after 200 shots with different objects: a hair (a), a wire  of 18
\textrm{$\mu$}m diameter (d) and a spider thread (g). Reconstruction of the holograms without iteration (b, e, h) and
after 100 iterations (c, f, i).} \label{results}
\end{figure}

\begin{figure}
\begin{center}
\includegraphics[width=8cm]{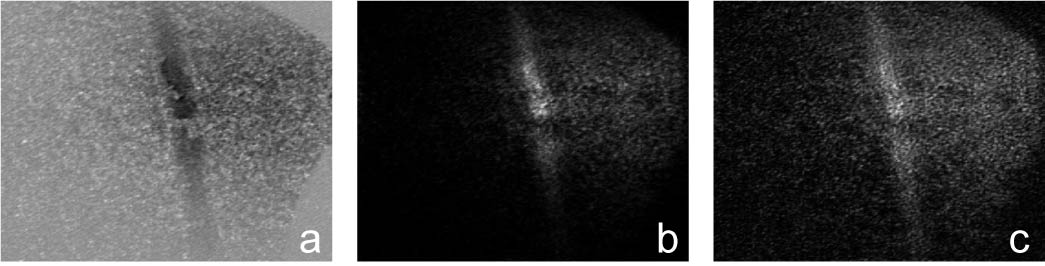}
\end{center}
\caption {Hologram recorded in a single shot of a wire of 18 \textrm{$\mu$}m diameter in (a). Reconstruction of the hologram without iteration (b) and
after 100 iterations (c).} \label{single_shot}
\end{figure}

The recorded holograms are reconstructed using the algorithm described above and are presented in Fig. \ref{results}. We observe that the reconstruction is possible for all the objects and that the phase retrieval process is efficient. It is possible to record and reconstruct holograms of wires of 6 $\mu$m diameter, which is still far above the theoretical resolution of our experimental setup (250 nm).
By comparison a single shot image of a 18 $\mu$m wide wire is shown in Fig. \ref{single_shot}. The reconstruction is still possible but with reduced quality, due to the low signal-to-noise ratio.

\section{Conclusion}
Holograms of thin wires are recorded using high order harmonics at 38 nm. A good spatial resolution is achieved with the help of a Schwarzschild objective. Since single shot holograms are recorded and reconstructed, we are able to reach the femtosecond temporal resolution range. A reconstruction program for digital in-line holography has been developed, tested and implemented. An iteration process was also introduced to improve the phase retrieval process of the reconstruction. With the help of this program we are able to reconstruct different objects.

This work proves that digital in-line holography using XUV radiation is a
possible alternative to microscopy and a step in its
optimization has been taken. Although there are many improvement
that can - and should - be done, this is a step towards
successfully performing time-resolved holography with extreme
ultraviolet radiation. The main improvements that will be made are mostly
experimental. It seems important to select only one harmonic order to
have a hologram which can be entirely reconstructed. This could be
possible by adding other multilayer optics efficiently selecting 
one harmonic order. Simultaneously, the harmonic photon flux should be improved by at least a factor of 10 to increase the signal-to-noise ratio. Promising techniques using, for example, improved phase matching \cite{BartelsNature2000,PfeiferAPB2005} or multicolor fields \cite{MauritssonPRL2006,KimPRL2005}
 make this within reach. In the future the possibility of single-shot measurements gives this technique a great potential for imaging femtosecond processes.

This work was supported by the Swedish Research Council, the Knut and Alice Wallenberg foundation, the Swedish Foundation for International Cooperation in Research and Higher Education (STINT) and the European STREP project TUIXS (Tabletop Ultra Intense XUV Sources).

\end{document}